\newcommand{\al}{\alpha}
\newcommand{\pa}{\partial}
\newcommand{\ep}{\epsilon}
\newcommand{\si}{\sigma}
\newcommand{\Si}{\Sigma}
\newcommand{\tsi}{\tilde \si}
\newcommand{\tet}{\tilde \eta}
\newcommand{\la}{\lambda}
\newcommand{\ta}{\tau}

\newcommand{\om}{\omega}

\newcommand{\De}{\Delta}

\newcommand{\rar}{\rightarrow}
\newcommand{\lrar}{\leftrightarrow}

\documentstyle[11pt,oneside,amssymb,array,amstex]{amsart}

\def\fun#1#2{\lower3.6pt\vbox{\baselineskip0pt\lineskip.9pt
  \ialign{$\mathsurround=0pt#1\hfil##\hfil$\crcr#2\crcr\sim\crcr}}}
\newskip\humongous \humongous=0pt plus 1000pt minus 1000pt

\newif\ifdtup

\newcommand{\non}{\nonumber}

\begin{document}

\begin{titlepage}

\begin{flushright}
Minneapolis TPI-MINN-98/04 \\ M\'exico ICN-UNAM 98-02\\solv-int/98005003
\\April 1998
\end{flushright}

\vskip 1.6cm

\begin{center}

{\Large Hidden Algebra of Three-Body Integrable Systems}

\vskip 0.6cm

{\it Alexander Turbiner}$^{\dagger}$
\vskip 0.5cm
Theoretical Physics Institute, University of Minnesota, Minneapolis,
\\MN 55455, USA\\ and\\
Instituto de Ciencias Nucleares, UNAM, Apartado Postal 70--543, \\04510
Mexico D.F., Mexico

\end{center}

\vskip 2 cm

\centerline{Abstract}

\begin{quote}
It is shown that all 3-body quantal integrable systems that emerge in the
Hamiltonian reduction method possess the same hidden algebraic structure.
All of them are given by a second degree polynomial in generators of an
infinite-dimensional Lie algebra of differential operators.
\end{quote}

\vfill
\noindent
$^\dagger$On leave of absence from the Institute for Theoretical
and Experimental Physics,
Moscow 117259, Russia\\
E-mail: turbiner@@tpi1.hep.umn.edu

\hskip 1.cm turbiner@@xochitl.nuclecu.unam.mx

\end{titlepage}

The Hamiltonian reduction method (see, for example, \cite{Olshanetsky:1983})
has provided several few-parameter families of integrable many-body 
potentials. The goal of this Letter is to show that all three-body,
rational and trigonometric integrable $A_2,BC_2, G_2$  Hamiltonians 
possess the {\it same} hidden algebra.
They are given by a second degree polynomial in generators of some
infinite-dimensional algebra. The information about their Hamiltonian 
reduction origin and the coupling constants is coded in the coefficients 
of this polynomial.
\vskip .5cm

{\bf 1.} {\it $A_2$ integrable system (rational case)}.

The Hamiltonian of the three-body Calogero model or, in other words, the 
rational $A_2$ integrable model is defined by
\begin{equation}
\label{e1.1}
        {\cal H}_{{\rm Cal}} = \frac{1}{2}\sum_{i=1}^{3}
\bigg[-\frac{\pa^{2}}{\pa x_{i}^{2}} + \om^2 x_{i}^{2}\bigg] +
g\sum_{i<j}^{3}\frac{1}{(x_{i}-x_{j})^{2}}\ ,
\end{equation}
where $g=\nu(\nu -1) > -\frac{1}{4}$ is the coupling constant and
$\om$ is the harmonic oscillator frequency. This Hamiltonian describes a
system of three identical particles on the line with pairwise interaction.
The ground state eigenfunction is given by
\begin{equation}
\label{e1.2}
\Psi_{0}^{(c)}(x) = \De^{\nu}(x) e^{-\om\frac{X^{2}}{2}}\ ,
\end{equation}
where $\De(x) = \prod_{i<j}|x_{i}-x_{j}|$ is the Vandermonde determinant
and $X^{2} = \sum_{i}x_{i}^{2}$. The Hamiltonian (\ref{e1.1}) is
$Z_2$-invariant, $x \rar -x$, which leads to two families of eigenstates:
even and odd. Throughout the paper we will deal with {\it even} eigenstates 
only. The odd eigenstates can be treated similarly and nothing conceptually new 
appears.

Basically, the internal dynamics of the system is
defined by relative motion. To study this relative motion let us
introduce the center-of-mass coordinate $Y\ =\ \sum_{j=1}^3 x_j$ and
the translation-invariant relative coordinates (Perelomov coordinates)
\cite{Perelomov:1971},
\begin{equation}
\label{e1.3}
y_i\ =\ x_i - \frac{1}{3} Y\ ,\quad i=1,2,3 \quad ,
\end{equation}
which obey the constraint $y_1+y_2+y_3=0$. To incorporate permutation
symmetry and translational invariance we consider the coordinates
\cite{Ruhl:1995}
\begin{equation}
\label{e1.4}
\ta_2 \ = \ -y_1^2-y_2^2-y_1y_2\ ,\
\ta_3 \ = \ -y_1y_2(y_1+y_2)\ ,
\end{equation}
or \cite{Rosenbaum:1997},
\begin{eqnarray}
\label{e1.5}
\la_1  =   \ta_2\ ,\qquad\qquad\la_2  =  \ta_3^2\ .
\end{eqnarray}
It is worth mentioning that the coordinates (\ref{e1.5}) are $Z_2$ symmetrical,
$\la_{1,2}(-x)=\la_{1,2}(x)$.
Performing a gauge rotation of the Hamiltonian (\ref{e1.1}),
\begin{equation}
\label{e1.6}
h_{\rm Cal}  = -2(\Psi_0^{(c)}(x))^{-1}{\cal H}_{\rm Cal}\Psi_0^{(c)}(x) \ ,
\end{equation}
and rewriting the resulting operator in the coordinates (\ref{e1.5}), we get
the following differential operator with polynomial coefficients,
\[
h_{\rm Cal}(\la_1,\la_2) = -2\la_1\pa^2_{\la_1\la_1}
        -12\la_2\pa^2_{\la_1\la_2}
        +{8\over 3}\la_1^2\la_2\pa^2_{\la_2\la_2} -
\]
\begin{equation}
\label{e1.7}
        -[4\om\la_1+2(1+3\nu)]\pa_{\la_1}
        -\bigl(12\om \la_2-{4\over 3}\la_1^2\bigr) \pa_{\la_2}\ .
\end{equation}
It can be called the {\it algebraic} form of $A_2$ Calogero model.

Following the philosophy of (quasi)-exact-solvability (see
\cite{Turbiner:1994}) let us try to find the hidden algebra of (\ref{e1.1})
as an origin of solvability of this model. It can be shown that
the operator (\ref{e1.7}) can be rewritten in terms of the generators of some
infinite-dimensional Lie algebra of differential operators generated
by the eight operators
\begin{xalignat}{2}
\label{e1.8}
L^1 & =  \pa_{\la_1}\qquad (-1,0)\ , &
L^2 & = \la_1\pa_{\la_1} - {n\over 3}\qquad (0,0)\ ,\non \\
L^3 &  =  2\la_2\pa_{\la_2}- {n\over 3}\quad (0,0)\ , &
& L^4 = \la_1^2\pa_{\la_1}+2\la_1\la_2
\pa_{\la_2} - n\la_1 \quad (+1,0)\ , \non \\
L^5 & = \pa_{\la_2}\qquad (0,-1)\ , &
L^6 & = \la_1\pa_{\la_2}\qquad (+1,-1)\ , \non \\
L^7 & =  \la_1^2\pa_{\la_2}\quad (+2,-1)\ , &
T & = \la_2\pa_{\la_1\la_1}^2\qquad (-2,+1) \ ,
\end{xalignat}
where the numbers in brackets $(a_1,a_2)$ mean the grading of the generator,
\linebreak
$A: \la_1^{k_1}\la_2^{k_2} \mapsto \la_1^{k_1+a_1}\la_2^{k_2+a_2}$.
This algebra was introduced at the first time in \cite{Rosenbaum:1997} and
was called there $g^{(2)}$.
The first seven generators $L_i$ form the $gl_2 \ltimes R^3$-algebra.
If $n$ is a non-negative integer number, the finite-dimensional irreducible
representation (\ref{e1.8}) appears with the invariant subspace
\begin{equation}
\label{e1.9}
W_n=(\la_1^{n_1}\la_2^{n_2} | 0\leq (n_1+2 n_2) \leq n) \ .
\end{equation}
The generators of the $gl_2 \ltimes R^3$-algebra, $L_1,\ldots L_7,$
act on $W_n$ reducibly, having the invariant subspace
\begin{equation}
\label{e1.10}
\tilde W_n=\langle \la_1^{p} | 0\leq p \leq n\rangle \ .
\end{equation}
It is worth mentioning that at $n=0$ the $gl_2 \ltimes R^3$-algebra becomes
the algebra of vector fields, which act on 2-Hirzebruch surface, $\Si_2$ and
the modules are the sections of holomorphic line bundles over this surface
(see \cite{ghko} and references therein).

Finally, when the operator (\ref{e1.7}) is written in terms of the generators
(\ref{e1.8}) it becomes,
\begin{equation}
\label{e1.11}
h_{\rm Cal}=-2L^2L^1-6L^3L^1+{4\over 3}L^7L^3
        -2(1+3\nu)L^1-4\om L^2-6\om L^3-{4\over 3}L^7\ ,
\end{equation}
where the parameter $n$ is equal to zero, $n=0$. This is called the {\it Lie-algebraic} form of the Calogero model.
The representation (\ref{e1.11}) contains the generators of the Borel 
subalgebra of $gl_2 \ltimes R^3$-algebra only, preserving the infinite 
flag of finite-dimensional representation spaces $W_n$ as well as the 
infinite flag of $\tilde W_n$, which proves the exact-solvability of 
the $A_2$ Calogero model.
It is worth noting that when the configuration space of (\ref{e1.1}) is
parametrized by the $\ta$-coordinates (\ref{e1.4})
the gauge-rotated Hamiltonian (\ref{e1.6}) can be rewritten in terms of the
generators of the $gl(3)$-algebra \cite{Ruhl:1995}. So the Calogero model
possesses two different hidden algebras, $g(2)$ and $gl(3)$ acting on the
configuration space in two different parametrizations.

\vskip .5cm

{\bf 2.} {\it $A_2$ integrable system (trigonometric case)}.

The Hamiltonian of the three-body Sutherland model, or in other words, the
trigonometric $A_2$ integrable model is defined by \cite{Olshanetsky:1983}
\begin{equation}
\label{e2.1}
{\cal H}_{\rm Suth} =
 -\frac{1}{2}\sum_{k=1}^{3}\frac{\pa^{2}}{\pa x_{k}^{2}}
 + \frac{g\al^2}{4}\sum_{k<l}^{3}\frac{1}{\sin^{2}(\frac{\al}{2}(x_{k} -
 x_{l}))} \ ,
\end{equation}
where $g=\nu (\nu-1) > -\frac{1}{4}$ is the coupling constant. The
ground state of this Hamiltonian is
\begin{equation}
\label{e2.2}
\Psi_{0}^{({\rm Suth})}(x) = (\De^{(trig)} (x))^{\nu}\ ,
\end{equation}
where $\De^{(trig)} (x) = \prod_{i<j}^3|\sin\frac{\al}{2}(x_{i}-x_{j})|$
is the trigonometric analog of the Vandermonde determinant.

To exhibit the dynamics of the system, we can introduce
the translation-invariant, permutation-symmetric, periodic
coordinates, either \cite{Ruhl:1995}
\[
\eta_2  =  \frac{1}{\al^2}[\cos(\al y_1)+\cos(\al
y_2)+\cos(\al (y_1+y_2))-3]\ ,
\]
\begin{equation}
\label{e2.3}
\eta_3  =  \frac{2}{\al^3}[\sin(\al y_1)+\sin(\al
y_2)-\sin(\al (y_1 + y_2))]\ ,
\end{equation}
or \cite{Rosenbaum:1997}
\begin{equation}
\label{e2.4}
\si_1  =  \eta_2\ ,\qquad \si_2  =  \eta_3^2\ .
\end{equation}
In the limit $\al \rar 0$, these coordinates become (4) or (5), respectively.
The coordinates (\ref{e2.4}) are symmetric with respect to $x \rar -x$,
$\si_{1,2}(-x) = \si_{1,2}(x)$.

Performing the gauge rotation of the Hamiltonian (\ref{e2.1}) 
(see (\ref{e1.6})) and extracting the center-of-mass motion, we get the 
{\it algebraic} form of the Hamiltonian of the Suthurland model
\[
h_{\rm Suth}  =
        -(2\si_1+{\al^2\over 2}\si_1^2-{\al^4\over
        24}\si_2) \pa_{\si_1\si_1}^2
        -(12+{8\al^2\over 3}\si_1)\si_2 \pa_{\si_1\si_2}^2
\]
\[
  +({8\over 3}\si_1^2\si_2-2\al^2\si_2^2) \pa_{\si_2\si_2}^2
        -\bigl[2(1+3\nu)+2(\nu+{1\over
        3})\al^2\si_1\bigr] \pa_{\si_1}
\]
\begin{equation}
\label{e2.5}
 +\biggl[{4\over 3}\si_1^2-({7\over 3}+4\nu )\al^2\si_2\biggr]
        \pa_{\si_2}\ ,
\end{equation}
(cf. (\ref{e1.7})).
This operator can be represented in terms of the generators
of the algebra $g^{(2)}$ with $n=0$ and $\la_1,\la_2$ replaced by
$\si_1,\si_2$, respectively,
\[
h_{\rm Suth}  =  -2 L^2 L^1 - 12 L^3 L^1 + {8\over 3} L^7 L^3 -
\al^2 (\frac{L^2 L^2}{2} + \frac{8 L^3 L^2}{3}+2 L^3 L^3)
\]
\begin{equation}
\label{e2.6}
  + \frac{\al^4}{24} T -2(1+3\nu) L^1 - {4\over 3}L^7
        - 2\bigl(\nu+\frac{1}{12}\bigr)\al^2L^2
        -4\bigl(\nu+\frac{1}{12}\bigr)\al^2L^3 \ ,
\end{equation}
This is the Lie-algebraic form of the Sutherland Hamiltonian. It contains
the generators of the Borel subalgebra of $gl_2 \ltimes R^3$-algebra and 
as well as the generator $T$.
In the limit $\al \rar 0$ the operator (\ref{e2.6}) coincides with the
operator (\ref{e1.11}) at $\om=0$.

It is worth noting that in the $\eta$-coordinates (\ref{e2.3})
the gauge-rotated Sutherland Hamiltonian (\ref{e2.5}) can be rewritten 
in terms of the generators of the $gl(3)$-algebra \cite{Ruhl:1995}.
So, the Sutherland Hamiltonian similarly to the Calogero one possesses 
two different hidden algebras: 
a $g^{(2)}$ algebra acting on the configuration space parametrized by
$\si$-coordinates and a $gl(3)$-algebra acting on the configuration space
parametrized by $\eta$-coordinates.
\vskip .5cm

{\bf 3.} {\it $BC_2$ integrable system (rational case)}.

The Hamiltonians of the $BC_{2}, B_{2}, C_{2}$ rational models do coincide and
they are given by (see \cite{Olshanetsky:1983})
\[
{\cal H}^{(r)}_{\rm BC_2} =-\frac{1}{2}(\frac{\pa^{2}}{\pa x_{1}^{2}} +
\frac{\pa^{2}}{\pa x_{2}^{2}}) +
\frac{\om^{2}}{2}(x_{1}^{2}+x_{2}^{2}) +
\]
\begin{equation}
\label{e3.1}
 g\left[ \frac{1}{(x_{1}-x_{2})^{2}} +
\frac{1}{(x_{1}+x_{2})^{2}} \right] +
\frac{g_{2}}{2}\left(\frac{1}{x_{1}^{2}} +\frac{1}{x_{2}^{2}}\right)\ ,
\end{equation}
where $g = \nu(\nu - 1)$ and $g_{2} = \nu_{2}(\nu_{2} - 1)$.
When the coupling constant $g_{2}$ tends to zero the Hamiltonian
${\cal H}^{(r)}_{\rm BC_2}$ degenerates to the Hamiltonian of the $D_{2}$
rational model. This Hamiltonian can be treated as the Hamiltonian of the
relative motion of some 3-body problem with non-identical particles (for
a discussion see \cite{Rosenbaum:1997}).

The ground state eigenfunction of the Hamiltonian (\ref{e3.1}) is given by
\begin{equation}
\label{e3.2}
\Psi_{0} = |x_{1}^2-x_{2}^2|^{\nu}|x_{1}x_{2}|^{\nu_{2}}
e^{-\frac{\om}{2}(x_{1}^{2}+x_{2}^{2})}\ ,
\end{equation}
(cf.(\ref{e1.2})). In order to encode the permutation symmetry
$x_i \lrar x_j$ and the reflection symmetry $x_i \rar -x_i$ of the
Hamiltonian (\ref{e3.1}), we introduce the coordinates \cite{Brink:1997}
\begin{equation}
\label{e3.3}
\tilde \si_1 = x_1^2 + x_2^2\ ,\ \tilde \si_2 = x_1^2 x_2^2\ .
\end{equation}

Now we perform the gauge rotation of (\ref{e3.1}) with ground state
eigenfunction (\ref{e3.2}) (see (\ref{e1.6})). Eventually, in the 
$\tilde \si$ coordinates the gauge-rotated $BC_2$ rational Hamiltonian 
takes its {\it algebraic} form,
\[
-h^{(r)}_{BC_2}(\tsi_1,\tsi_2)=4\tsi_1\pa^2_{\tsi_1\tsi_1}
+16 \tsi_2 \pa^2_{\tsi_1 \tsi_2}+4\tsi_1\tsi_2
\pa^2_{\tsi_2\tsi_2}
\]
\begin{equation}
\label{e3.4}
+4[(1+\nu_2+2\nu)-\om\tsi_1]\pa_{\tsi_1}+
2[(1+\nu_2)\tsi_1-4\om\tsi_2]\pa_{\tsi_2} \ ,
\end{equation}
This operator can be rewritten in terms of the $gl_2 \ltimes R^3$-algebra
generators as
\newpage
\[
-h^{(r)}_{BC_2}\ =\  4L^2L^1 + 8L^1L^3 + 2L^3L^6 \ +
\]
\begin{equation}
\label{e3.5}
4(1+\nu_2+2\nu)L^1-
4\om L^2+2(1+\nu_2)L^6-4\om L^3 \ ,
\end{equation}
which is the Lie-algebraic form of $BC_2$ rational model. It contains 
the generators of the Borel subalgebra of $gl_2 \ltimes R^3$-algebra only.
The eigenvalues of $h^{(r)}_{BC_2}$ are given by
\[
\ep_{n,k}\ =\ 4\om(2n-k)\ ,\ n=0,1,2,\ldots\ ,\ k=0,1,2,\ldots n\ .
\]
It is worth noting that unlike the $A_2$ cases (see Sections 1, 2)
the gauge-rotated Hamiltonian (\ref{e3.1}) in $\tilde \si$-coordinates 
allows the representation either in terms of the generators of the 
$g^{(2)}$ algebra or the generators of the $gl(3)$-algebra \cite{Brink:1997}.

\vskip .5cm

{\bf 4.} {\it $BC_2$ integrable system (trigonometric case)}.

The Hamiltonians for $B_{2}, C_{2}$ and $D_{2}$ trigonometric models are
special cases of the general $BC_2$ Hamiltonian \cite{Olshanetsky:1983}
\[
{\cal H}^{(t)}_{BC_2} = -\frac{1}{2}(\frac{\pa^{2}}{\pa x_{1}^{2}} +
\frac{\pa^{2}}{\pa x_{2}^{2}})+ \frac{g}{4}\left[
\frac{1}{\sin^{2}(\frac{1}{2}(x_{1}-x_{2}))} +
\frac{1}{\sin^{2}(\frac{1}{2}(x_{1}+x_{2}))} \right] +
\]
\begin{equation}
\label{e4.1}
\frac{g_{2}}{4}\sum_{i=1}^{2}\frac{1}{\sin^{2}(x_{i})} +
\frac{g_{3}}{8}\sum_{i=1}^{2}\frac{1}{\sin^{2}(\frac{x_{i}}{2})}
\end{equation}
where $g = \nu(\nu - 1)$, $g_{2} = \nu_{2}(\nu_{2} - 1)$ and $g_{3} =
\nu_{3}(\nu_{3} + 2\nu_{2} - 1)$.
From the general Hamiltonian the $B_{2}$, $C_{2}$ and $D_{2}$ cases
are obtained as follows:
\begin{itemize}
\item $B_{2}$ case:  $\nu_{2}=0$,
\item $C_{2}$ case:  $\nu_{3}=0$,
\item $D_{2}$ case:  $\nu_{2}=\nu_{3}=0$.
\end{itemize}

The Hamiltonian (\ref{e4.1}) can be treated as the Hamiltonian of the
relative motion of a three-body problem with non-identical particles 
(see \cite{Rosenbaum:1997}).

The ground state wave function is \cite{Olshanetsky:1983,Bernard:1995}
\begin{equation}
\label{e4.2}
\Psi_{0} =
|\sin(\frac{1}{2}(x_{1}-x_{2}))|^{\nu}
|\sin(\frac{1}{2}(x_{1}+x_{2}))|^{\nu}\prod_{i=1}^{2}
|\sin(x_{i})|^{\nu_{2}}|\sin(\frac{x_{i}}{2})|^{\nu_{3}} \ .
\end{equation}

In order to reveal both the permutation and reflection symmetry and 
the periodicity of (\ref{e4.1}), let us introduce the coordinates
\begin{equation*}
\eta_1 (\al)  =  \cos \al x_1 + \cos \al x_2 \ ,\
\eta_2 (\al)  =  \cos \al x_1 \cos \al x_2  \ ,
\end{equation*}
and a modification of the them
\begin{equation}
\label{e4.3}
\tet_1 \ = \ \frac{4}{\al^2}-\frac{2}{\al^2} \eta_1 (\al)\ ,\
\tet_2 \ = \ \frac{4}{\al^4}-\frac{4}{\al^4} [\eta_1 (\al)-\eta_2(\al)]\ ,
\end{equation}
In the limit $\al \rar 0$ the $\tet$'s became the $\tilde \si$'s
(see (\ref{e3.3})). Performing a gauge rotation of the Hamiltonian 
(\ref{e4.1}) with the ground state eigenfunction (\ref{e4.2}) 
(see (\ref{e1.6})) and then changing the coordinates to the $\tet$ 
coordinates, we get the algebraic form of the $BC_2$ trigonometric 
Hamiltonian,
\[
-h^{(t)}_{BC_2}(\tet_1,\tet_2)=(4\tet_1-\al^2\tet_1^2+2\al^2\tet_2)
\pa^2_{\tet_1\tet_1}
+2\tet_2(8-\al^2\tet_1) \pa^2_{\tet_1 \tet_2}\ +
\]
\[
2\tet_2(2\tet_1-\al^2\tet_2)\pa^2_{\tet_2\tet_2} +
4[\nu_2+\nu_3+2\nu-\frac{\al^2}{4}(\nu_2+\frac{\nu_3}{2}+\nu)\tet_1]
\pa_{\tet_1} \ +
\]
\begin{equation}
\label{e4.4}
2[(\nu_2+\nu_3)\tet_1-\al^2(\nu_2+\frac{\nu_3}{2}+\nu)\tet_2]\pa_{\tet_2}\ .
\end{equation}
If $\nu_3=1$ and $\al \rar 0$ we arrive at (\ref{e3.4}). The operator
(\ref{e4.4}) can be rewritten in terms of the $g^{(2)}$-algebra generators, 
\[
h^{(t)}_{BC_2}=-4L^2L^1 + \al^2L^2 L^2 -2\al^2 T-8L^1L^3+\al^2 L^3 L^2-
2L^3L^6\ -
\]
\[
4(\nu_3+\nu_2+2\nu)L^1 +
\frac{\al^2}{4}(\nu_2+\frac{\nu_3}{2}+\nu-4)L^2 +
\]
\begin{equation}
\label{e4.5}
\frac{\al^2}{2}(\nu_2+\frac{\nu_3}{2}+\nu) L^3 - 2(\nu_3+\nu_2)L^6\ ,
\end{equation}
which is the Lie-algebraic form of the $BC_2$ trigonometric Hamiltonian.

The eigenvalues of $h^{(t)}_{BC_2}$ are given by
\[
\ep_{n,k}\ =\ \al^2[nk+2(n-k)^2 +(\nu_2+\frac{\nu_3}{2}+\nu)(2n-k)]\ ,
\]
\begin{equation}
\label{e4.6}
n=0,1,2,\ldots\ ,\ k=0,1,2,\ldots n\ .
\end{equation}
The limit $\al \rar 0$ makes no sense since there are no polynomial
eigenfunctions of the operator (\ref{e4.4}) in this limit.

It is worth noting that unlike the $A_2$ models
the gauge-rotated Hamiltonian (\ref{e4.4}) in $\tet$-coordinates can be
rewritten in terms of the generators of the $g^{(2)}$ algebra as well
as the $gl(3)$-algebra \cite{Brink:1997}.

\vskip .5cm

{\bf 5.} {\it $G_2$ integrable system (rational case)}.

The rational $G_2$ Hamiltonian describes a three-body system of the identical
particles with two- and three-body interactions
\[
{\cal H}_{\rm G_2}^{(r)} =
 \frac{1}{2}\sum_{k=1}^{3}\bigg[-\frac{\pa^{2}}{\pa x_{k}^{2}}
+ \om^2 x_k^2 \bigg]\ +
\]
\begin{equation}
\label{e5.1}
 g\sum_{k<l}^{3}\frac{1}{(x_{k} - x_{l})^2}
 + g_1\sum\begin{Sb} k<l \\ k,l \neq m\end{Sb}^{3}
 \frac{1}{(x_{k} + x_{l}-2x_{m})^2}  \ ,
\end{equation}
where $g=\nu (\nu-1) > -\frac{1}{4}$ and $g_1=3\mu (\mu -1) > -\frac{3}{4}$
are the coupling constants associated with the two-body and three-body interactions, respectively. The ground-state eigenfunction is given by
\begin{equation}
\label{e5.2}
\Psi_{0}^{({\rm r})}(x) = (\De^{(r)}(x))^{\nu} (\De_1^{(r)}(x))^{\mu}
e^{-\frac{1}{2}\om \sum x_i^2} \ ,
\end{equation}
where $\De^{(r)}(x)=\prod_{i<j}^3|x_i-x_j|$ and
$\De_1^{(r)}(x)=\prod_{i<j; \ i,j\neq k}|x_i+x_j-2x_k|$.

The result of a gauge rotation of the Hamiltonian (\ref{e5.1}) with the
ground state eigenfunction (\ref{e5.2}) (see (\ref{e1.6})) can be written 
in terms of the coordinates $\la_1, \la_2$ given by (\ref{e1.5}). Thus,
\[
h_{\rm G_2}^{(r)}  =  -2\la_1\pa^2_{\la_1\la_1}
        -12\la_2\pa^2_{\la_1\la_2}
        +{8\over 3}\la_1^2\la_2\pa^2_{\la_2\la_2}
\]
\begin{equation}
\label{e5.3}
   -\bigg\{4\om\la_1+2[1+3(\mu+\nu)]\bigg\}\pa_{\la_1}
        -\bigl( 12\om\la_2-{4\over
        3}\la_1^2\bigr)\pa_{\la_2}\ ,
\end{equation}
(cf.(\ref{e1.7})). It is quite amazing  that the difference between 
(\ref{e1.7}) and (\ref{e5.3}) is only in replacement $\mu \rar (\mu+\nu)$. 
This is the {\it algebraic} form of the rational $G_2$ model, which  
admits a representation in terms of the generators of the algebra $g^{(2)}$ 
at $n=0$,
\[
h_{\rm G_2}^{(r)}  =  -2 L^2 L^1 - 12 L^3 L^1 + {8\over 3} L^7 L^3\ -
\]
\begin{equation}
\label{e5.4}
2[1+ 3(\mu+\nu)] L^1 - 4\om L^2 - 12\om L^3 -\frac{4}{3} L^7\ ,
\end{equation}
(cf. (\ref{e1.11})). Equation (\ref{e5.4}) is the $g^{(2)}$
{\it Lie-algebraic} form of the rational $G_2$ model. This form depends on
the generators of the $gl_2 \ltimes R^3$-subalgebra only. This implies
that (\ref{e5.4}) possesses {\it two} invariant subspaces, $W_n$ and
$\tilde W_n$ (see (\ref{e1.9})--(\ref{e1.10})). Thus we are led to the
conclusion that there exists a family of eigenfunctions depending only on
the variable $\la_1$.
In fact this property was already known both for the present model
\cite{Wolfes:1974} as well as for the general many-body Calogero model.
It was used in \cite{Minzoni:1996} to construct
quasi-exactly-solvable deformation of the general Calogero model.

It is worth noting that in the $\ta$-coordinates (\ref{e1.4})
the gauge-rotated $G_2$ rational Hamiltonian (\ref{e5.1}) can be rewritten in
terms of the generators of the $gl(3)$-algebra \cite{Rosenbaum:1997}.
Thus this Hamiltonian possesses two different hidden algebras:
a $g^{(2)}$ algebra acting on the configuration space parametrized by the
$\la$ coordinates and a $gl(3)$-algebra acting on the configuration space
parametrized by the $\ta$ coordinates.

\vskip .5cm

{\bf 6.} {\it $G_2$ integrable system (trigonometric case)}.

The Hamiltonian for the trigonometric $G_2$ model has the form
\[
{\cal H}_{\rm G_2} =
 -\frac{1}{2}\sum_{k=1}^{3}\frac{\pa^{2}}{\pa x_{k}^{2}}
 + \frac{g}{4}\sum_{k<l}^{3}\frac{1}{\sin^{2}(\frac{1}{2}(x_{k} - x_{l}))}
\]
\begin{equation}
\label{e6.1}
 + \frac{g_1}{4}\sum\begin{Sb} k<l \\ k,l \neq m\end{Sb}^{3}
 \frac{1}{\sin^{2}(\frac{1}{2}(x_{k} + x_{l}-2x_{m}))}\ ,
\end{equation}
where $g=\nu (\nu-1) > -\frac{1}{4}$ and $g_1=3\mu (\mu -1) > -\frac{3}{4}$
are the coupling constants associated with the two-body and three-body interactions, respectively. From the physical point of view (\ref{e6.1}) describes a system of three identical particles. The ground-state 
eigenfunction is given by
\begin{equation}
\label{e6.2}
\Psi_{0}^{(t)}(x) = (\De^{(trig)}(x))^{\nu} (\De_1^{(trig)}(x))^{\mu}\ ,
\end{equation}
where $\De^{(trig)}(x),\ \De_1^{(trig)}(x)$ are the trigonometric
analogies of the Vandermonde determinant and are defined by
\[
\De^{(trig)} (x) = \prod_{i<j}^3 |\sin\frac{1} {2}(x_{i}-x_{j})| \ ,
\]
\[
\De_1^{(trig)} (x) = \prod\begin{Sb} k<l \\ k,l \neq m\end{Sb}^3|\sin\frac{1}
{2}(x_{i}+x_{j}-2x_{k})| \ .
\]

Let us introduce the permutation-symmetric, translation-invariant,
periodic coordinates:
\begin{eqnarray}
\label{e6.3}
\tsi_1 & = & {1\over\al^2}\biggl[\cos(\al
(y_1-y_2))+\cos(\al (y_2-y_3))+\cos(\al (y_3-y_1))-3\biggr]\ ,
\nonumber \\
\tsi_2 & = & {4\over\al^6}\biggl[
\sin(\al (y_1-y_2))+\sin(\al (y_2-y_3))+\sin(\al (y_3-y_1))\biggr]^2 \ ,
\end{eqnarray}
In these coordinates the trigonometric $G_2$
Hamiltonian, gauge-rotated with the ground state eigenfunction (\ref{e6.2})
(see (\ref{e1.6}) with the factor $(\frac{2}{3})$ instead of 2) becomes
\[
h_{G_2}^{(t)} \ = \
-(2\tsi_1+{\al^2\over 2}\tsi_1^2-{\al^4\over 24}\tsi_2)\pa_{\tsi_1\tsi_1}^2-
(12+{8\al^2\over 3}\tsi_1)\tsi_2\pa_{\tsi_1\tsi_2}^2\ +
\]
\[
+({8\over 3}\tsi_1^2\tsi_2-2\al^2\tsi_2^2)\pa_{\tsi_2\tsi_2}^2
- \bigl\{2[1+3(\mu +2\nu)]+{2\over 3}(1 + 3\mu +
       4 \nu)\al^2\tsi_1\bigr\}\pa_{\tsi_1}\ +
\]
\begin{equation}
\label{e6.4}
\bigg\{{4\over 3}(1+4\nu)\tsi_1^2-[{7\over 3}
        +4(\mu +\nu )]\al^2\tsi_2\bigg\}
                \pa_{\tsi_2}\ .
\end{equation}
This is the
{\it algebraic} form of the trigonometric $G_2$ model and $h_{G_2}^{(t)}$
can be written in terms of the generators of the algebra $g^{(2)}$ 
containing both the generators of $gl_2 \ltimes R^3$ and the generator $T$.
The explicit expression is given by
\[
h_{G_2}^{(t)}\ =\ -2 L^2 L^1-6 L^3 L^1+{4\over 3} L^7 L^3
-\frac{\al^2}{6} (3L^2 L^2+8 L^3 L^2+ 3L^3 L^3) + \frac{\al^4}{24}T
\]
\[
-2[1+3(\mu +2\nu )] L^1 - {4\over 3}(1-4\nu) L^7
        - (2\mu + \frac{8}{3}\nu+\frac{1}{6})\al^2 L^2
\]
\begin{equation}
\label{e6.5}
 -\bigl[\frac{1}{6} + 2(\mu +\nu )\bigr]\al^2 L^3 \ .
\end{equation}

The operator (\ref{e6.5}) can be easily reduced to triangular form
by introducing the new variables
\begin{equation}
\label{e6.6}
\rho_1  =  \tsi_1 \ , \qquad\qquad
\rho_2  =  \tsi_2 +\frac{4}{\al^2}\tsi_1^2 \ .
\end{equation}
It is worth noticing that if $\al \rar 0$ this change of variables becomes
singular, reflecting the non-existence of bound states for
the Calogero and $G_2$ rational models in the absence of the harmonic
oscillator term in the potential, $\om=0$.

In new coordinates the Hamiltonian takes the form
\[
h_{G_2}^{(t)}\ =\ -(2\rho_1 + \frac{2}{3}\al^2\rho_1^2 -
\frac{\al^4}{24}\rho_2)\pa_{\rho_1\rho_1}^2  -
(12\rho_2 + 2\al^2\rho_1\rho_2
-\frac{16}{\al^2}\rho_1^2)\pa_{\rho_1\rho_2}^2\ -
\]
\[
(2\al^2\rho_2^2 + \frac{96}{\al^2}\rho_1\rho_2 -
\frac{256}{\al^4}\rho_1^3 )\pa_{\rho_2\rho_2}^2 -
[2(1+3\mu+6\nu) +
\frac{2}{3}(1+3\mu+4\nu)\al^2\rho_1]\pa_{\rho_1}\ -
\]
\begin{equation}
\label{e6.7}
\{2(1+2\mu+2\nu)\al^2\rho_2 +
\frac{16}{\al^2}(2+3\mu+6\nu)\rho_1\}\pa_{\rho_2} \ ,
\end{equation}
and it is easy to check that it is indeed a triangular operator. Evidently
this operator can be rewritten in terms of the $g^{(2)}$-generators as
\begin{eqnarray}
\label{e6.8}
h_{G_2}^{(t)} & = & -2 L^2 L^1 - 6 L^1 L^3 + \frac{\al^4}{24}T
- \frac{2}{3}\al^2 L^2 L^2 - \al^2 L^2 L^3 - \frac{\al^2}{2} L^3 L^3 +
\nonumber \\
& \ & \frac{16}{\al^2}L^7L^1 - \frac{48}{\al^2}L^3L^6 +
\frac{256}{\al^2}L^6L^7 - 2(1 + 3\mu + 6\nu)L^1 -
\nonumber \\
& \ &  (2\mu + \frac{8}{3}\nu)\al^2L^2 - 2(\mu +\nu)\al^2 L^3 -
\frac{16}{\al^2}(2+3\mu+6\nu)L^6 \ .
\end{eqnarray}

Using either of the above representations, the energy levels of the
Hamiltonian $H_{G_2}$ can be easily found and are given by
\begin{equation}
\label{e6.9}
-{\cal E}_{n,m}  = \biggl[(n-2m-1)(n+m+1)+3n\mu+2(2n-m)\nu+1\biggr]\al^2 \ ,
\end{equation}
where $n$ and $m$ are quantum numbers,
\begin{equation}
n=0,1,2,3,\ldots \ , \qquad\qquad  0\leq m\leq\bigl[\frac{n}{2}\bigr] \ .
\end{equation}
We would like to emphasize that unlike the rational and trigonometric 
$A_2, BC_2$ and rational $G_2$ Hamiltonians, the trigonometric $G_2$ 
Hamiltonian has the only one hidden algebra $g^{(2)}$.

As a general conclusion we would like to stress that all three-body 
integrable quantal systems originating from the Hamiltonian reduction 
method, but with arbitrary coupling constants, possess the same hidden 
algebra $g^{(2)}$. Generically, the orthogonal polynomials in two variables
associated the eigenfunctions of $A_2-BC_2-G_2$ integrable models do not
appear in Krall-Sheffer classification scheme of 2d orthogonal polynomials
(see, for example, \cite{l}).

\section*{Acknowledgements}

The author thanks the Theoretical Physics Institute, University of
Minnesota for its kind hospitality. Useful discussions
with M.~Rosenbaum on the early stage of the work and also with
M.~Shifman are highly appreciated. The work is partially supported by
DOE grant DE-FG02-94ER40823.

\def\href#1#2{#2}

\begingroup\raggedright\endgroup


\begin{thebibliography}{10}


\bibitem{Olshanetsky:1983}
M.~A. Olshanetsky and A.~M. Perelomov, ``Quantum integrable systems related to
  Lie algebras,'' {\em Phys. Rep.} {\bf 94} (1983) 313.

\bibitem{Perelomov:1971}
A.~M. Perelomov, ``Algebraic approach to the solution of one-dimensional
  model of $N$ interacting particles,'' {\em Teor. Mat. Fiz.} {\bf 6} (1971)
  364 (in Russian); English translation: {\it Sov. Phys. -- Theor. and Math.
Phys. \bf  6} (1971) 263.

\bibitem{Ruhl:1995}
W.~R{\"u}hl and A.~Turbiner, ``Exact solvability of the Calogero and Sutherland
  models,'' {\em Mod. Phys. Lett.} {\bf A10} (1995) 2213--2222,
  \href{http://xxx.lanl.gov/abs/hep-th/9506105}{{\tt hep-th/9506105}}.

\bibitem{Rosenbaum:1997}
M.~Rosenbaum, A.~Turbiner and A.~Capella,
``Solvability of the $G_2$ integrable system,"
{\em Intern. Journ. Mod. Phys. \bf A} (1998) to appear,
\href{http://xxx.lanl.gov/abs/solv-int/9707005}{{\tt solv-int/9707005}}.

\bibitem{Turbiner:1994}
A.~V. Turbiner, ``Lie algebras and linear operators with invariant subspace,''
  in {\em Lie algebras, cohomologies and new findings in quantum mechanics}
  (N.~Kamran and P.~J. Olver, eds.), AMS, vol.~160, pp.~263--310, 1994;\\
 ``Lie-algebras and Quasi-exactly-solvable Differential Equations'',
    in {\em CRC Handbook of Lie Group Analysis of Differential Equations},
    Vol.3: New Trends in Theoretical Developments and Computational
    Methods, Chapter 12, CRC Press (N.~Ibragimov, ed.), pp.~331-366, 1995;
             {\tt hep-th/9409068}.

\bibitem{ghko}
       A. Gonz\'alez-Lop\'ez, J. Hurtubise, N. Kamran and P.J. Olver,
       ``Quantification de la cohomologie des alg\`ebres de Lie de champs
       de vecteurs et fibr\'es en droites sur des surfaces complexes compactes",
       {\em C.R.Acad.Sci.(Paris), S\'erie I \bf 316} (1993) 1307-1312

\bibitem{Brink:1997}
    L.~Brink, A.~Turbiner and N.~Wyllard, ``Hidden Algebras of the (super) Calogero
     and Sutherland models,'' Preprint ITP 97-05 and ICN-UNAM/97-02
    \href{http://xxx.lanl.gov/abs/hep-th/9705219}{{\tt hep-th/9705219}}
    {\em Journ. Math. Phys.\bf 39}(1998) 1285-1315

\bibitem{Bernard:1995}
       D.~Bernard, V.~Pasquier and D.~Serban, ``Exact solution of long-range
      interacting spin chains with boundaries,'' {\em Europhys. Lett. \bf 30}
      (1995) 301-305, {\tt hep-th/9501044}

\bibitem{Wolfes:1974}
J.~Wolfes, ``On the three-body linear problem with three-body interaction,"
             {\em J. Math. Phys. \bf 15} (1974) 1420-1424.

\bibitem{Minzoni:1996}
A.~Minzoni, M.~Rosenbaum and A.~Turbiner,
``Quasi-Exactly-Solvable Many-Body Problems,"
{\em Mod. Phys. Lett. \bf A11} (1996) 1977-1984,
\href{http://xxx.lanl.gov/abs/hep-th/9606092}{{\tt hep-th/9606092}}.

\bibitem{l}
       L.L.~Littlejohn, ``Orthogonal polynomial solutions to
       ordinary and partial differential equations",
       Proceedings of an International Symposium on Orthogonal Polynomials
       and their Applications, Segovia, Spain, Sept.22-27, 1986,
       Lecture Notes in Mathematics No.1329, M.Alfaro et al. (Eds.),
       Springer-Verlag (1988), pp. 98-124.


\end{thebibliography}
\end{document}